\documentclass[a4paper,11pt,onecolumn,accepted=2021-07-14]{quantumarticle}
\pdfoutput=1

\usepackage[italicdiff]{physics}
\usepackage[numbers,sort]{natbib}
\usepackage{hyperref}
\hypersetup{colorlinks=true,
allcolors=quantumviolet}

\usepackage[utf8]{inputenc}
\usepackage[english]{babel}
\usepackage{amsmath}
\usepackage{amssymb}
\usepackage{lipsum}
\usepackage{appendix}
\usepackage{float}
\usepackage{caption}
\usepackage{subcaption}
\usepackage{enumitem}
\usepackage{doi}

\begin{document}

\title{Nonclassical trajectories in head-on collisions}

\author[1,2]{A Kumar}\orcid{0000-0003-3639-6468}\email{akumar18@ph.iitr.ac.in}
\author[2]{T Krisnanda}\orcid{0000-0002-6360-5627}
\author[1]{P Arumugam}\orcid{0000-0001-9624-8024}
\author[2,3,4]{T Paterek}\orcid{0000-0002-8490-3156}
\affil[1]{Department of Physics, Indian Institute of Technology Roorkee, Roorkee 247667, India}
\affil[2]{School of Physical and Mathematical Sciences \\ Nanyang Technological University, Singapore 637371, Singapore}
\affil[3]{Institute of Theoretical Physics and Astrophysics \\ Faculty of Mathematics, Physics and Informatics, University of Gda\'{n}sk, 80-308 Gda\'{n}sk, Poland}
\affil[4]{MajuLab, International Joint Research Unit UMI 3654 \\ CNRS, Universit\'{e} C\^{o}te d’Azur, Sorbonne Universit\'{e}, \\ National University of Singapore, Nanyang Technological University}

\maketitle

\begin{abstract}
Rutherford scattering is usually described by treating the projectile either classically or as quantum mechanical plane waves. Here we treat them as wave packets and study their head-on collisions with the stationary target nuclei. We simulate the quantum dynamics of this one-dimensional system and study deviations of the average quantum solution from the classical one. These deviations are traced back to the convexity properties of Coulomb potential. Finally, we sketch how these theoretical findings could be tested in experiments looking for the onset of nuclear reactions.
\end{abstract}

\section{Introduction}

The seminal theoretical discussion of scattering angles in the Rutherford experiment is based on the Coulomb interaction between point charges modeling the alpha projectiles and the stationary gold nuclei~\cite{RutherfordExp_Students,RutherfordExp_Prof}.
In the conventional quantum approach, the projectiles are described as incident plane waves, and the collision is studied in asymptotic limits under suitable approximations~\cite{book_CJJoachain}. A fuller approach that we pursue here is to compute the time dependence of the quantum evolution. With this in mind we study the simplest case of Rutherford scattering, i.e., the head-on collision.

As in the original discussion we model the nuclei as stationary sources of the Coulomb potential, but in contradistinction we describe the alpha particles by incident Gaussian wave packets. This leads to several predictions that differ from their classical counterparts. We focus on the average quantum behavior and show that it does not recover the classical solutions. In particular, the quantum projectile does not approach the target as close as its classical counterpart, the quantum dynamics is not symmetric about the time of collision, and the expected quantum trajectories do not overlap with the classical ones even in the asymptotic limits after the collision.

We trace back these discrepancies to the Ehrenfest theorem and emphasize that the average quantum dynamics recovers classical solutions only for potentials that are at most quadratic in the position operator~\cite{ZP45.455,book-QMech-BCHall,book-QMech-MaxJammer}.
We derive inequalities between average quantum and classical forces that hold for any cubic potential as well as for potentials with fixed convexity properties.
This is directly applicable to Coulomb or gravitational forces and clearly shows that average quantum dynamics are different from their classical counterparts in a plethora of physically interesting scenarios.

Finally, we briefly discuss tunneling through the Coulomb barrier to infer the distance of closest approach and its dependence on the initial spread of the incident Gaussian wave packet.
The tunneling probability is an experimentally accessible parameter as
the particles that have crossed the barrier give rise to nuclear reactions~\cite{ZP-54.656,Nature-106.14}. In addition to computing the probability in the dynamical quantum model we also show the conditions under which the Wentzel-Kramers-Brillouin formula accurately approximates it~\cite{PR-40.621}.
In particular, we note that the latter may be orders of magnitude off for low energy projectiles even for a negligible momentum dispersion (see also Refs.~\cite{PLA-220.41,PLA-225.303,JPA-37.2423,PLA-378.1071}).


\section{Ehrenfest theorem and classical limit}

Before moving to the collisions we would like to present a general discussion on a comparison between the average Schr\"odinger dynamics and the classical one, especially that some textbooks give an incorrect statement that the average quantum trajectory recovers the classical motion~\cite{book_QM_Eisberg}.
This comes in relation to the Ehrenfest theorem showing that the quantum equations of motion for average position and momentum:
\begin{equation}
\dv{t}\ev{x} = \frac{\ev{p}}{m}, \hspace{5mm} \dv{t}\ev{p} = -\ev{V'(x)},
\label{eq:ehrenfesteqs}
\end{equation}
have the same general form as the Hamilton's equations of classical mechanics:
\begin{equation}
\dv{t}x = \frac{p}{m}, \hspace{5mm} \dv{t}p = -V'(x),
\label{eq:hamiltoneqs}
\end{equation}
where $x$ is the position, $p$ is the momentum, $m$ is the mass of the particle and $V'$ stands for the gradient of potential to which we will refer as (negative) force. In what follows, we only consider the one-dimensional motion.

While the general form of ~\eqref{eq:ehrenfesteqs} and ~\eqref{eq:hamiltoneqs} is the same, they become identical only if the average of the force equals the force for the average position~\cite{book-QMech-BCHall}:
\begin{equation}
\ev{V'(x)} = V'(\ev{x}).
\label{eq:EeqH}
\end{equation}
For arbitrary wave functions, this condition is satisfied only for potentials that are at most quadratic in $x$.
Already a cubic potential shows that there is a consistent difference between the quantum and the classical forces.
Namely, the derivative of the potential $V'(x) = a_1 + a_2 \, x + a_3 \, x^2$. Assuming $a_i > 0$, the non-negativity of variance implies $\ev{V'(x)} \geq V'(\ev{x})$. Similar inequalities follow for derivatives of convex or concave potentials using Jensen's inequalities~\cite{ActaMath-30.175}. Many potentials of natural interactions have well-defined convexity properties, e.g., the repulsive Coulomb potential we study here. In this case, the wave packet's average quantum force can be very different from what a classical particle experiences in the same potential. This leads to consistent differences between the average quantum trajectory and the classical one.

\section{Classical trajectories}

For comparison, we first show how to compute the classical trajectories.
Consider a projectile, initially at $x = x_0$ (negative), is shot towards the target, located at $x = 0$.
We follow the tradition that projectiles are propagating from left to right. If the particle's initial kinetic energy is $T_0$, the law of conservation of energy implies $T(x) + V(x) = T_0 + V(x_0) \equiv E_0$,
with the Coulomb potential given by
\begin{equation}
V(x) = Z_{P}Z_{T} \alpha \hbar c / | x |,
\label{CoulombPotential}
\end{equation}
where $Z_{P}$ and $Z_{T}$ are the atomic numbers of the projectile and the target, respectively, and $\alpha$ is the fine structure constant.
Accordingly, the classical equation of motion is
\begin{equation}
\dv{x}{t} = \pm \sqrt{\frac{2E_0}{m}} \sqrt{1-\frac{d_{\mathrm{cl}}}{|x|}},
\label{cleom}
\end{equation}
where $d_{\mathrm{cl}} = Z_P Z_T \alpha \hbar c / E_0$ is the classical distance of closest approach.
The above equation can be integrated and gives the following trajectories:
\begin{eqnarray}
 \frac{d_\mathrm{cl}}{2} \ln \left( \frac{1+\sqrt{\frac{x+d_\mathrm{cl}}{x}}}{1-\sqrt{\frac{x+d_\mathrm{cl}}{x}}} \right) - x\sqrt{\frac{x+d_\mathrm{cl}}{x}} = \sqrt{\frac{2E_0}{m}}(t-\tau_{\mathrm{cl}}) \ \text{sign}(t-\tau_{\mathrm{cl}}),
\label{xcl}
\end{eqnarray}
where $\tau_{\mathrm{cl}}$ denotes the time of classical collision:
\begin{equation}
\tau_{\mathrm{cl}} = \sqrt{\frac{m}{2E_0}} \left[ \frac{d_\mathrm{cl}}{2} \ln \left( \frac{1+\sqrt{\frac{x_0+d_\mathrm{cl}}{x_0}}}{1-\sqrt{\frac{x_0+d_\mathrm{cl}}{x_0}}} \right) - x_0\sqrt{\frac{x_0+d_\mathrm{cl}}{x_0}}\right].
\label{taucl}
\end{equation}
Since Eq.~\eqref{xcl} is transcendental, we solve for $x(t)$ numerically.


\section{Quantum trajectories}
\label{sec:formalism}

\begin{figure}[!b]
\centering
\includegraphics[width=0.5\linewidth]{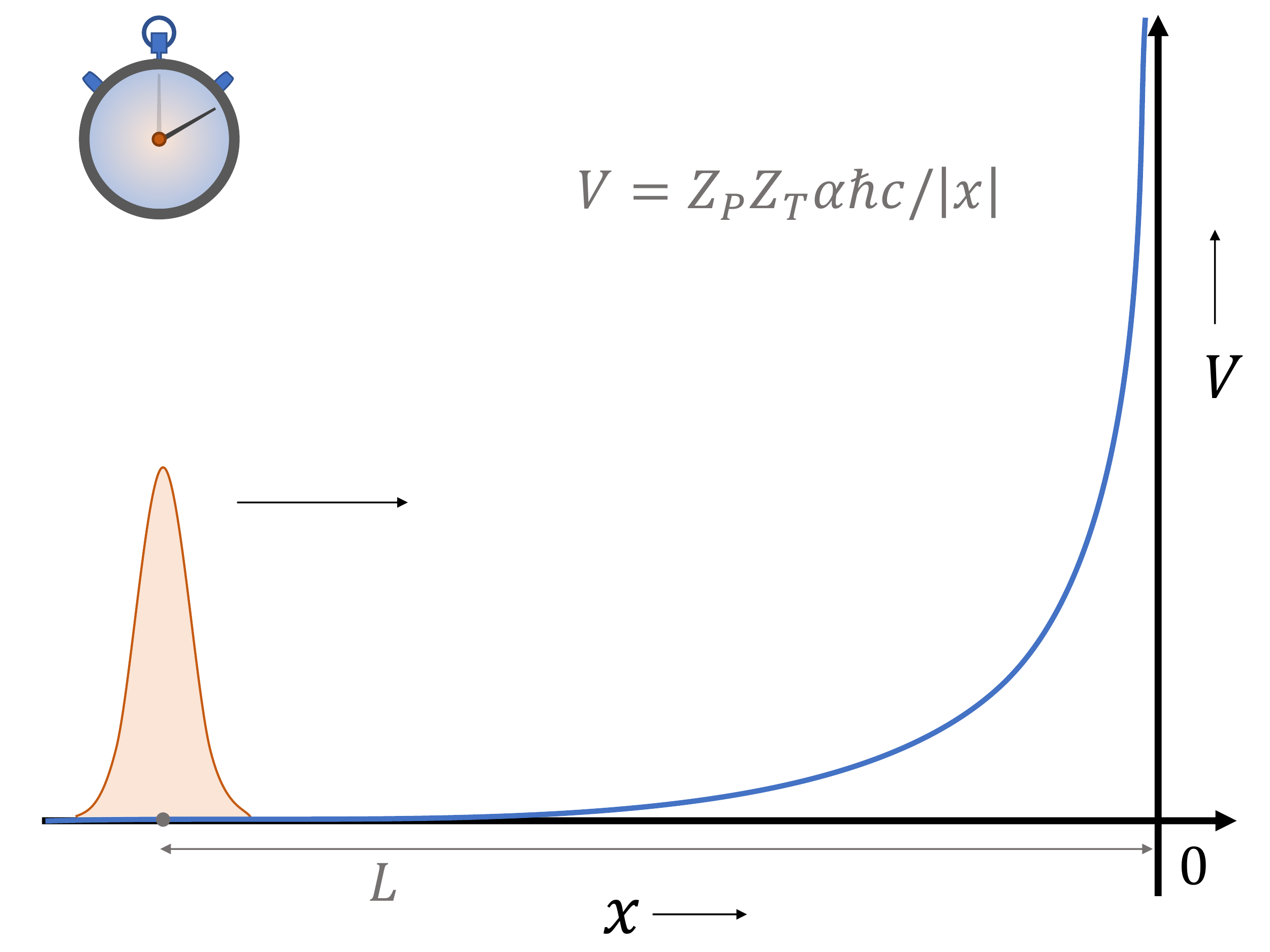}
\caption{Schematic of the experiment. The projectiles are prepared as Gaussian wave packets, which evolve in the Coulomb potential of the target nucleus fixed at the origin. The symbols have their usual meanings and are defined explicitly in Sec~\ref{sec:formalism}.}
\label{FIG_SCHEME}
\end{figure}

The quantum setup is shown schematically in Fig.~\ref{FIG_SCHEME}.
We describe here the methods used to obtain the average quantum dynamics of the projectile.

In non-relativistic quantum theory, the properties of a particle are described by a wave function $\psi(x,t)$, which evolves in accordance with the time-dependent Schr\"{o}dinger equation. For time-independent interactions, one can alternatively use the evolution operator $ U$ to write
\begin{equation}
\psi(x,t+\Delta t) = U(t+\Delta t,t) \ \psi(x,t) \equiv \exp(- i \frac{H \Delta t}{\hbar}) \ \psi(x,t),
\end{equation}
where $ H = -\frac{\hbar^2}{2m}\pdv[2]{x} + V( x)$ is the Hamiltonian. Any truncation in a series expansion of $ U$ leads to loss of unitarity, and consequently a change in the norm of wave function over time. We circumvent this problem and calculate the time-development by implementing the Cayley's form of evolution operator~\cite{JCP-68.2794,JCP-134.041101,book_CompPhys_FJVesely,CPC-208.9}.
Since the wave function is expanding continuously during the collision event, the simulation becomes numerically demanding as time progresses. To circumvent this we have developed a method to dynamically allocate the size of discretization of space, details of which are given in Appendix~\ref{appendix:cayley}.

Finally, in the classical picture the projectile particles are modelled as point charges shot towards the target from a distance $L$ with a kinetic energy $T_0$. In our quantum simulations, we model the projectiles with Gaussian wave packets centered at the same distance $L$, with a width $\sigma$ and an average momentum $P=\sqrt{2mT_0}$. Therefore, the wave function describing the quantum state of the projectile at $t=0$ is given by ($x_0 = -L$, \ $p_0 = P$)
\begin{equation}
\psi(x,0) = \sqrt{\frac{1}{\sigma\sqrt{2\pi}}} \exp{-\frac{(x-x_0)^2}{4\sigma^2} + i\frac{p_0}{\hbar}(x-x_0)}.
\end{equation}
We now describe several parameters that show differences between the classical and the average quantum dynamics.


\section{Comparison of trajectories}

\begin{figure}[!t]
\centering

\begin{subfigure}{\textwidth}
\centering
\includegraphics[width=\linewidth]{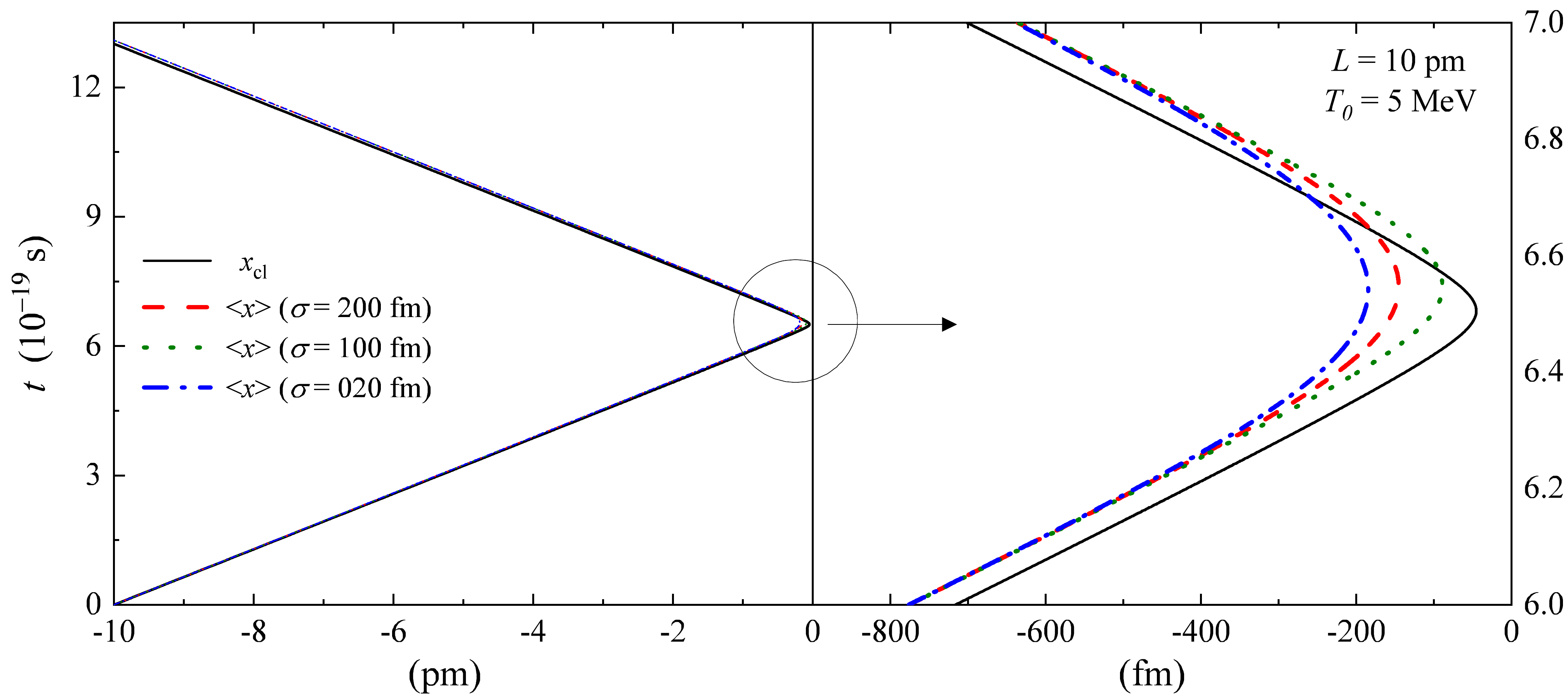}
\caption{Classical and quantum  trajectories. The right panel shows a magnified view of the scattering event. The quantum trajectories are the results from the dynamical simulations.}
\label{fig:Traj_10K5MeV}
\end{subfigure}

\vspace{5mm}

\begin{subfigure}{\textwidth}
\centering
\includegraphics[width=\linewidth]{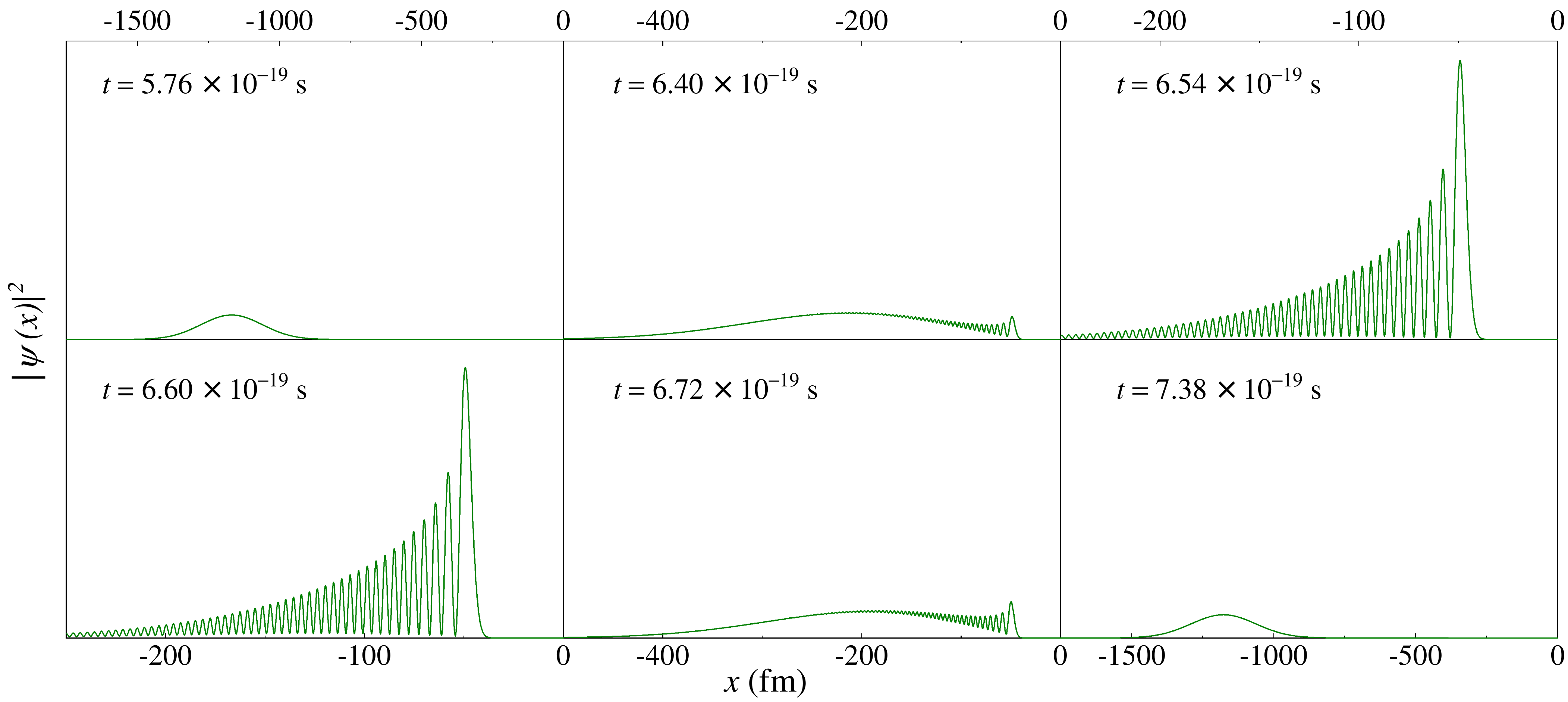}
\caption{Snapshots of the wave packet at different times for the case of $\sigma = 100$ fm.}
\label{fig:wavefunc_10K5MeV}
\end{subfigure}

\caption{Time-dependent properties of the alpha particle wave packets shot in the Coulomb potential of a gold nucleus fixed at the origin. The projectiles are shot from the distance $L = 10$ pm with the initial width of Gaussian wave packet $\sigma$ and average momentum $P=\sqrt{2mT_0}$, where $T_0 = 5$ MeV, and $m$ is the mass of an alpha particle. $x_\text{cl}$ represents the classical path, and $\ev{x}$ is the expected position of the wave packet. Numerical details are given in Appendix~\ref{appendix:constants}.}
\label{fig:10K5MeV}
\end{figure}

We follow the original Rutherford scattering experiment and study the collision of alpha particles with gold nuclei. Hence we set $Z_P=2$ and $Z_T=79$ throughout this work. Typical classical and quantum trajectories are presented in Fig.~\ref{fig:Traj_10K5MeV}.
In both cases $L$ = 10 pm and $T_0$ = 5 MeV, i.e., the typical energy of alpha particles in the original Rutherford scattering experiment.
We now systematically discuss the following quantities: distance of closest approach, the effects of a finite spread of the projectile's initial wave function, the time of collision, and the asymptotic behavior of trajectories.

\subsection{Distance of closest approach} \label{sec:doca}

As seen in Fig.~\ref{fig:Traj_10K5MeV} the quantum projectiles are, on average, reflected from a bigger distance to the target. This can be intuitively understood by invoking either convexity of the Coulomb interaction or the Heisenberg uncertainty principle. In the latter case, note that a quantum particle cannot be stopped completely. There is always some momentum dispersion which leads to a non-zero kinetic energy. It follows that the maximal potential energy cannot be as big as in the classical case, and hence the distance of the closest approach increases. In the former case, Jensen's inequality at time $t = 0$ allows us to write:
\begin{equation}
|\ev{F}| \sim \ev{\frac{1}{x^2(0)}} \geq \frac{1}{\ev{x(0)}^2} = \frac{1}{x_0^2} =  \frac{1}{x_{\mathrm{cl}}^2(0)} \sim |F_{\mathrm{cl}}|,
\label{inequality_teq0}
\end{equation}
Accordingly, the average quantum mechanical force is stronger (more repulsive) than the force experienced by the classical particle. As a result, the quantum projectile moves slower than its classical counterpart at the very beginning of its journey. We emphasize that this inequality holds at the beginning of the evolution, and at later times it may reverse (as we will show later).
Nevertheless, the quantum projectile never gets as close to the target as the classical one. In the next section we derive conditions for the quantum distance of closest approach to differ from its classical counterpart minimally.

Fig.\ref{fig:wavefunc_10K5MeV} shows the wave function at different times for the case of $\sigma = 100$ fm. As the wave packet approaches closer to the target, the leading edge of the wave packet is reflected back. This interferes with the trailing incident part to create rapid oscillations, which eventually die out as the wave packet travels back attaining a near-Gaussian shape~\cite{EJP-20.29,AJP-66.252,AJP-35.177}. As in the case of collision with a hard wall, the qualitative similarity between $\ev{x}$ and $x_\text{cl}$ is not evident when one looks at the wave packet itself~\cite{PhyScr-71.136}.

\subsection{Quantum distance of closest approach}

We define the quantum distance of closest approach as the smallest average position to the target:
\begin{equation}
d_{\mathrm{qm}} = \min(|\ev{x}|).
\end{equation}
As seen in Fig.~\ref{fig:Traj_10K5MeV} this quantity depends on the initial spread of the wave function of the projectile.
We now give physical arguments which determine the optimal initial spread $\sigma_0$, for which the difference $d_{\mathrm{qm}} - d_{\mathrm{cl}}$ is the smallest.

\begin{figure}[!t]
\centering
\includegraphics[width=0.5\linewidth]{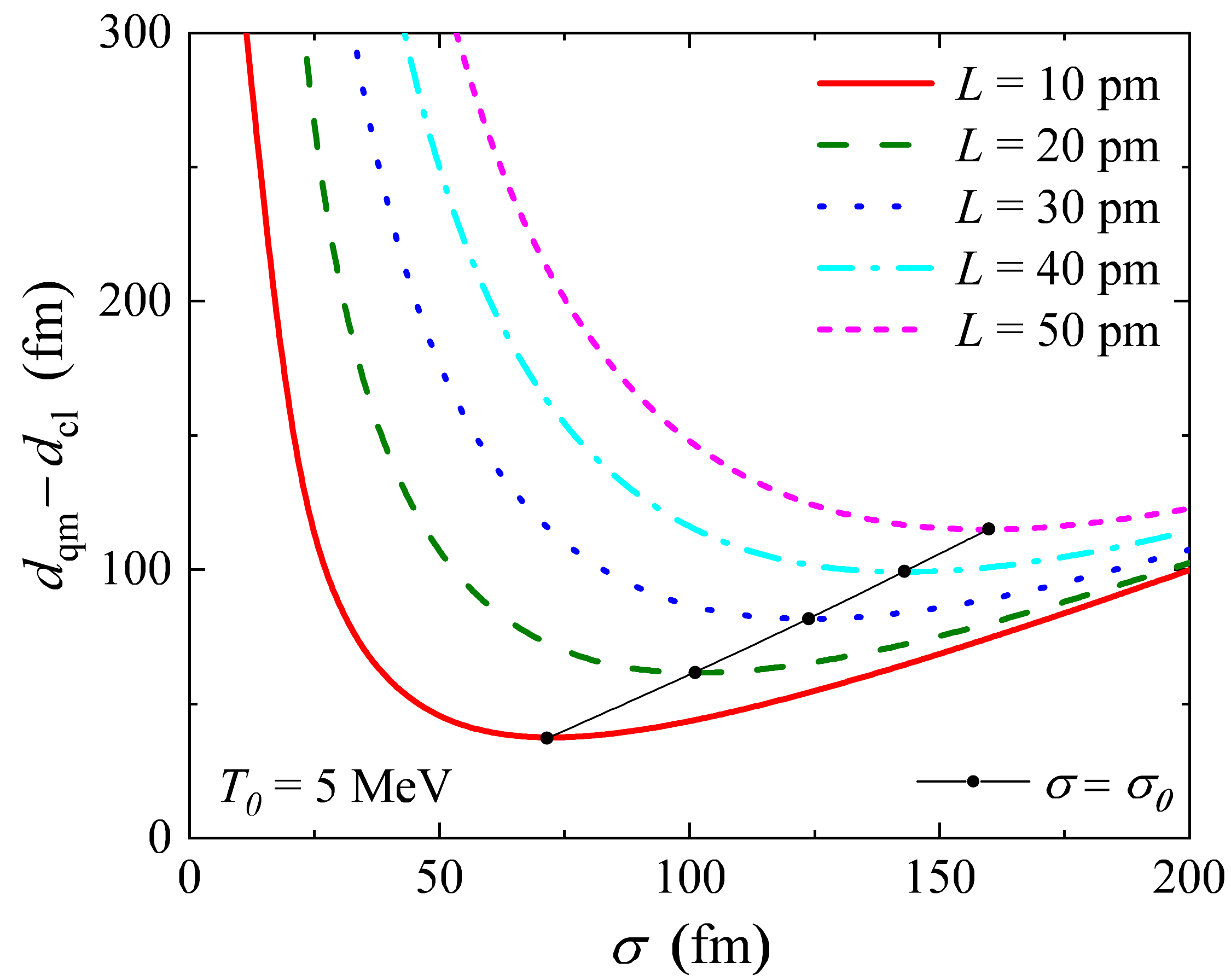}
\caption{Difference between the classical and quantum distances of closest approach. The numerically found minima of the difference are marked with dots for various launching distances. In the main text we argue that they are achieved for the initial width of the projectile given by $\sigma_0 = \sqrt{\hbar L/2P}$, where $P$ is the momentum of a classical alpha particle with kinetic energy $T_0$. Numerical details are given in Appendix~\ref{appendix:constants}.}
\label{fig:FindingSig0_5MeV}
\end{figure}

Since the projectile is launched from a large distance $L \gg  d_{\mathrm{cl}}$,
the time of collision satisfies $\tau_{\mathrm{cl}} \approx mL/P$.
We approximate the position spread before the collision by the value obtained for a free quantum evolution:
\begin{equation}
\sigma(\tau_{\mathrm{cl}}) = \sigma\sqrt{1+ \frac{\hbar^2 \tau_{\mathrm{cl}}^2}{4m^2\sigma^4}} = \sigma \sqrt{1+\left(\frac{\hbar L}{2P}\right)^2\frac{1}{\sigma^4}}.
\label{eq:sig0}
\end{equation}
We established in Eq.~\eqref{inequality_teq0} that the quantum wave packet feels a stronger force compared to its classical counterpart. As an implication the wave packet, on average, is reflected from a farther distance, i.e., $d_\text{qm} > d_\text{cl}$. For a given potential, the quantum averages at a given time are closer to their classical values for wave functions that are more and more concentrated in space. Therefore the difference $d_\text{qm} - d_\text{cl}$ is smaller for narrow wave functions, and it is minimal when $\sigma(\tau_{\mathrm{cl}})$ is minimal. Eq.~\eqref{eq:sig0} suggests that this happens for the initial spread given by
\begin{equation}
\sigma_0 = \sqrt{\frac{\hbar L}{2P}}
\label{EQ_IN_SPREAD}
\end{equation}
and corresponds to the spread at the collision time $\sigma(\tau_{\mathrm{cl}}) = \sqrt{2} \, \sigma_0$.
Fig.~\ref{fig:FindingSig0_5MeV} shows that this is in an excellent agreement with the numerically obtained minimas in $d_\text{qm}$.

\begin{figure*}[!t]
\centering
\includegraphics[width=\linewidth]{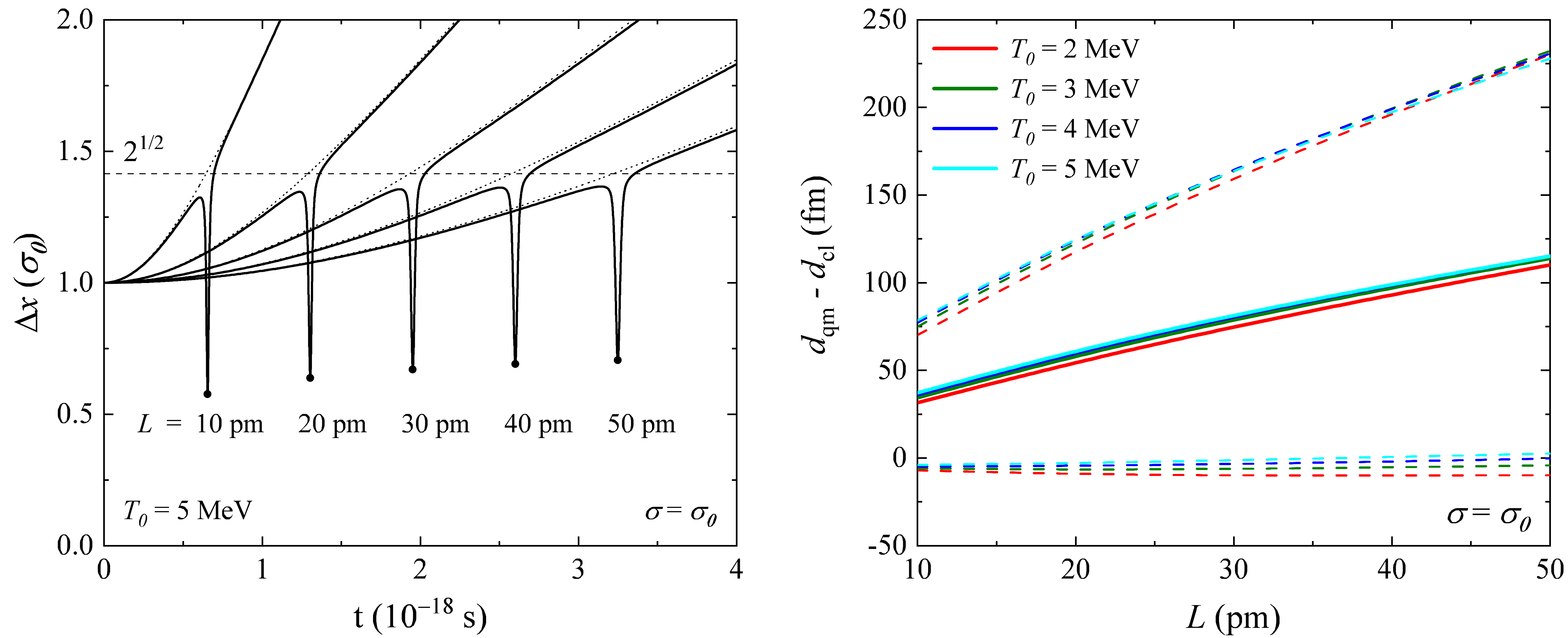}
\caption{Quantum distance of closest approach and the position spread during the collision. The initial width of wave packets is $\sigma_0 = \sqrt{\hbar L/2P}$, where $P$ is the momentum of a classical alpha particle with kinetic energy $T_0$. The left panel shows the position spreads of alpha particles with $T_0 = 5$ MeV for various launch distances $L$, with the dotted lines showing the spreads of a freely evolving particle. The black dots show the time of quantum collision, i.e. when the spread is minimum and the average position is $d_\text{qm}$. The right panel shows that the classical distance of closest approach is within a standard deviation (dashed lines) from the quantum distance of closest approach (the lower dashed line is close to zero). The energy dependence is negligible. Numerical details are given in Appendix~\ref{appendix:constants}.}
\label{fig:qdoca}
\end{figure*}

Furthermore, we establish the range accessible to the quantum distance of closest approach for the energies considered in this work. The left panel of Fig.~\ref{fig:qdoca} shows the position spread of wave packets launched from various distances.
The collisions correspond to the sudden drops of the position spread, i.e., the wave packet gets compressed when the projectile approaches the target. The distance of closest approach matches the minima in these curves, which are all smaller than the initial spread of Eq.~\eqref{EQ_IN_SPREAD}, i.e, the uncertainty in $d_\text{qm}$ is less than $\sigma_0$. Next, we show in the right panel of Fig.~\ref{fig:qdoca} that the classical distance of closest approach is within the range of one standard deviation from the quantum distance of closest approach. Finally, this translates to
\begin{equation}
d_{\mathrm{cl}} < d_{\mathrm{qm}}  < d_{\mathrm{cl}} + \sqrt{\frac{\hbar L}{2 P}}.
\end{equation}

\subsection{Origin of the optimal spread}

Mathematically, Eq.~\eqref{eq:EeqH} encoding the equivalence between the classical and Ehrenfest dynamics, is satisfied for any potential if the wave function is given by the Dirac delta.
The more the wave function is concentrated in space, the more similar classical and average quantum positions are.
Physically, however, due to the Heisenberg uncertainty principle, there is no evolution which preserves the Dirac delta, and we show that this leads to the optimal initial spread.
For the values presented in Fig.~\ref{fig:Traj_10K5MeV}, the optimal spread is about $50$ fm (see green dotted line).
If the initial spread is larger (compare with the red dashed line), the wave function spreads further and the deviation from the classical trajectory is higher. But for smaller initial spreads (compare with the blue dashed-dotted line), the free quantum evolution dominating initially predicts faster spreading due to large initial momentum dispersion.  Therefore, the standard deviation becomes larger by the time the wave packet reaches close to the target, and the deviation from the classical trajectory is again higher.

\subsection{Time asymmetry of quantum collision}

As already seen in Fig.~\ref{fig:Traj_10K5MeV}, the classical and quantum collisions happen at different times. The classical time $\tau_{\mathrm{cl}}$ is obtained by requiring vanishing momentum in Hamilton's solution.
The quantum collision time $\tau_{\mathrm{qm}}$ corresponds to vanishing average momentum in Ehrenfest's solution.
This time matches with the minima in the position spread shown in Fig.~\ref{fig:qdoca}.

\begin{figure}
\centering
\includegraphics[width=\linewidth]{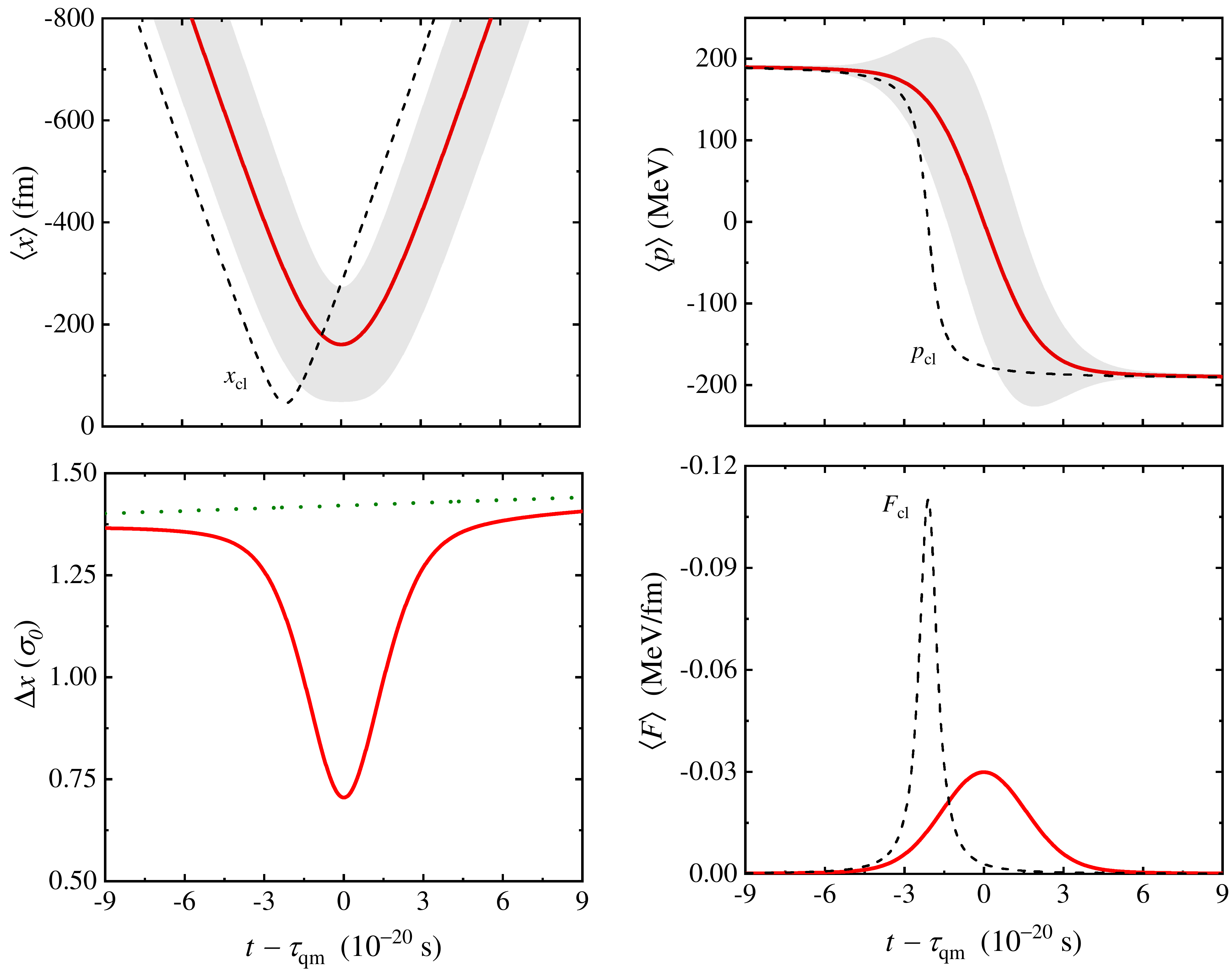}
\caption{Expected quantum properties (solid red lines) of an alpha particle about the quantum collision time. The dashed black lines present classical predictions. The dotted green line in the lower-left panel gives position spread for a freely evolving alpha particle wave packet (as if there was no Coulomb interaction). The gray shaded regions mark one standard deviation. The projectiles are launched from a distance $L = 50$ pm. The initial width of the wave packet is $\sigma_0 = \sqrt{\hbar L/2P}$, where $P$ is the momentum of a classical alpha particle with kinetic energy $T_0 = 5$ MeV. Numerical details are given in Appendix~\ref{appendix:constants}.}
\label{fig:MicroProp_50K5MeV}
\end{figure}

A distinguishing feature of the quantum trajectory is its asymmetry with respect to the collision time.
Recall that the classical trajectory is symmetric in time, i.e., $x_\text{cl}(\tau_{\mathrm{cl}} - \Delta t) = x_\text{cl}(\tau_{\mathrm{cl}} +\Delta t)$, and similarly for momentum $p_\text{cl}(\tau_{\mathrm{cl}} -\Delta t) = - p_\text{cl}(\tau_{\mathrm{cl}} + \Delta t)$. In particular, the projectile returns back to its original launch distance exactly at time $T = 2 \, \tau_{\mathrm{cl}}$. Our calculations show that the quantum projectile makes a collision at a different time $\tau_{\mathrm{qm}} (> \tau_{\mathrm{cl}})$, and the evolution for $t > \tau_{\mathrm{qm}}$ is not a mirror image of evolution for $t < \tau_{\mathrm{qm}}$. The return journey of a quantum particle takes a longer time than the onward one, and the wave packet returns at its launch position at time $T > 2 \, \tau_{\mathrm{qm}} (> 2 \, \tau_{\mathrm{cl}})$.
For the case in Fig.~\ref{fig:MicroProp_50K5MeV} ($\sigma = 159.85$ fm), the time taken for the return journey is longer by $\approx 1.3 \times 10^{-22}$ s. Such effects are more prominent for wave functions with a larger momentum variance, e.g., the time difference is $\approx 1.1 \times 10^{-21}$ s for the case of $\sigma = 20$ fm in Fig.~\ref{fig:10K5MeV}. The rate of increase of position spread itself increases with variance in the initial momentum, and given that the potential has a well-defined convexity, this tends to amplify the asymmetries over time.

The discussed asymmetry in quantum trajectories is yet another manifestation of the convexity in Coulomb potential. Consider a point at distance $X$ from the target. Irrespective of whether it is traveling towards or away from the target, a classical particle feels the exact same force when it passes through that point, i.e. $-V'(X)$. Accordingly, the Hamilton's solution of Eq.~\eqref{eq:hamiltoneqs} has to be symmetric about $\tau_{\mathrm{cl}}$. The quantum case is much more interesting; Fig.~\ref{fig:qdoca} and~\ref{fig:MicroProp_50K5MeV} show that the position spread immediately after the collision is larger than what it was before. This means that the width of the wave packet, when it passes through point $X$, is larger during its return journey. As a consequence, the average force experienced by the quantum projectile at that point, i.e. $-\ev{V'(X)}$, is different in the onward and the return journeys. Accordingly the Ehrenfest's solution of Eq.~\eqref{eq:ehrenfesteqs} has to be asymmetric about $\tau_{\mathrm{qm}}$. For the parameters used in this work, the classical trajectory is not even statistically within the quantum prediction, i.e. $x_{\text{cl}}$ and $p_{\text{cl}}$ are outside $\ev{x} \pm \Delta x$ and $\ev{p} \pm \Delta p$ (see Fig.~\ref{fig:MicroProp_50K5MeV}).

\subsection{Classical and quantum force}

We started our discussion by showing that initially the quantum projectile is repelled more than the corresponding classical counterpart.
For example, alpha particle of $T_0$ = 5 MeV launched from $L$ = 50 pm feels an average force $\ev{F} \approx 1.000031 F_{\mathrm{cl}}$ at $t=0$.
Since for later times $\ev{x} < x_\text{cl}$ (negative, see Fig.~\ref{fig:Traj_10K5MeV}) we obtain:
\begin{equation}
|\ev{F}| \sim \ev{\frac{1}{x^2(0)}} \geq \frac{1}{\ev{x(t)}^2} <  \frac{1}{x_{\mathrm{cl}}^2(t)} \sim |F_{\mathrm{cl}}|.
\label{inequality_tgt0}
\end{equation}
The second inequality starts dominating closer to the target, and during the collision the classical force far exceeds the quantum one as seen in Fig.~\ref{fig:MicroProp_50K5MeV}.

\section{Proposed experiments}

\begin{figure}[!b]
\centering
\includegraphics[width=0.5\linewidth]{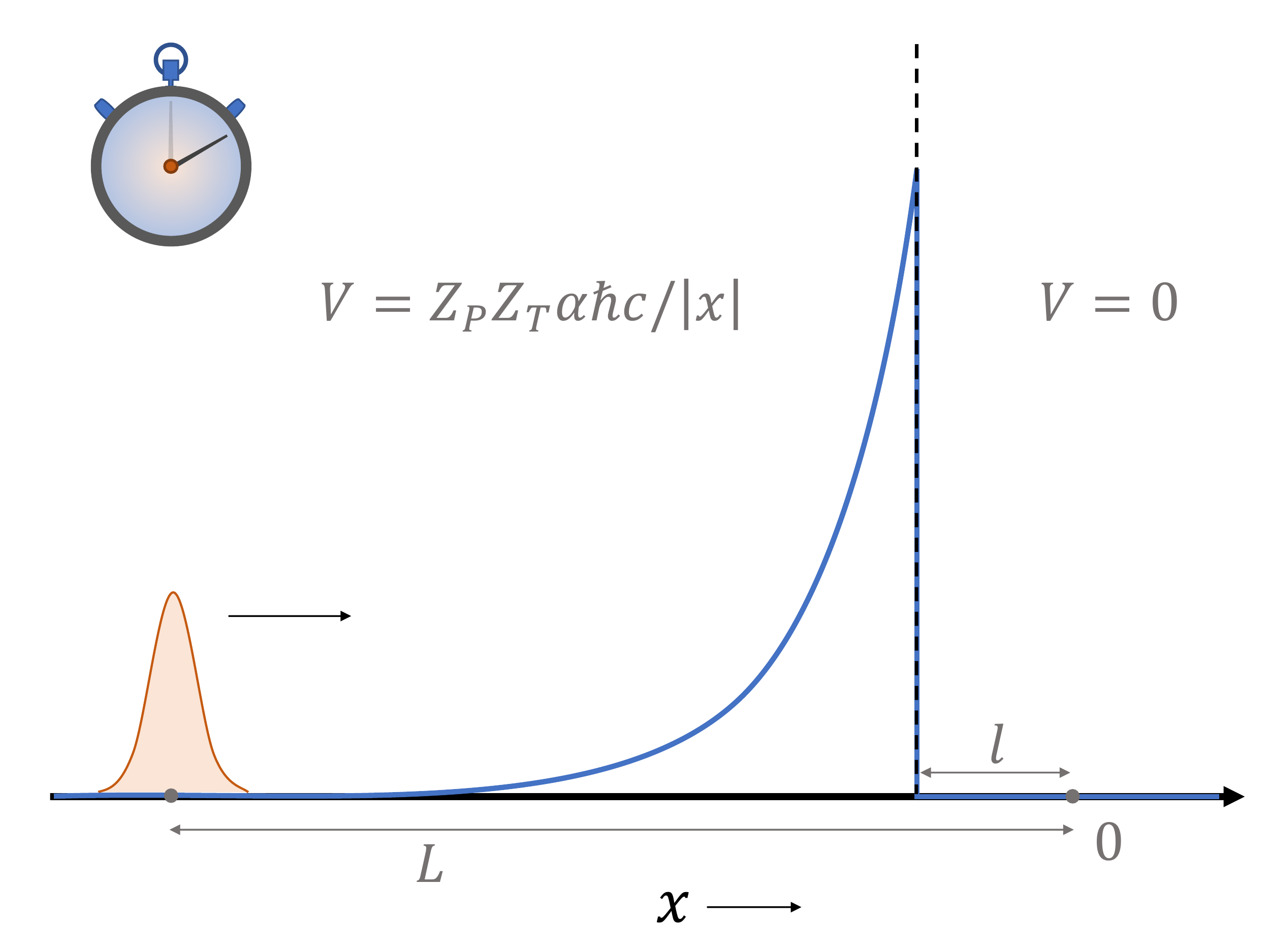}
\caption{Tunneling through the Coulomb barrier. The barrier is plotted in blue. The projectiles are prepared as Gaussian wave packets and shot from the left side. The size of the well, $l = 25$ fm, is the typical distance just outside the range of strong interaction.}
\label{FIG_BARRIER}
\end{figure}

Having established the differences between the classical and average quantum dynamics, we now move to a possibility of their experimental verification.

The basic idea is to look for the onset of nuclear reactions. We model the potential in the vicinity of the nucleus by the Coulomb barrier given in Fig.~\ref{FIG_BARRIER} and defined by:
\begin{equation}
V(x) =
\begin{cases}
Z_{P}Z_{T}\alpha \hbar c/|x|, \hspace{5mm} x \leq l, \\
0, \hspace{25.5mm} x > l.
\end{cases}
\label{barrier}
\end{equation}
In our simulations we varied the launching distance about $50$ pm,
and observe that the results are practically independent of $L$.

Under the classical model of the collision, if the projectile can cross the barrier it can be observed as a nuclear reaction. In such a case, the classical distance of closest approach is smaller than the size where the barrier is truncated. We take as a well-established fact that alpha particles cannot be prepared with only one momentum value and repeated momentum measurement on the projectile before the collision returns a normal distribution.
For this reason, within the classical model, we deal with an ensemble of projectiles with Gaussian momentum probability density.
Accordingly, they cross the Coulomb barrier only probabilistically with the corresponding probability $P_{\mathrm{cl}}$ plotted by the dashed line in Fig.~\ref{fig:TunnProb_50K}.
Conversely, from the measured probability of nuclear reaction, one can estimate the variance in position and momentum of the classical ensemble and compute the mean distance of closest approach.

Within the quantum model, the initial randomness is encoded by the Gaussian wave function. We compute its evolution in the presence of the Coulomb barrier and determine the probability that the particle tunnels through:
\begin{equation}
P_T = \lim\limits_{t \rightarrow \infty} \int_{l}^{\infty} dx \ |\psi(x,t)|^2.
\end{equation}
The result is plotted as a solid curve in the left panel of Fig.~\ref{fig:TunnProb_50K} and it is considerably different from the classical result for initial wave functions with position spread bigger than about $10$ fm.
For smaller position spreads the quantum and classical curves are approaching each other. Finally, for Dirac delta position distribution, in both cases we have flat momentum distribution, and hence half of the particles tunnel through.
The range of very small position spreads is not accessible in our numerical calculations because the wave function spreads very fast such that the dynamics of approaching the barrier already consumes all of the computational resources. We again emphasize that the nuclear reaction cross-section directly translates to the spread of the projectile's initial wave function. This provides an interesting way of experimentally determining this spread.

Note also that simultaneous measurement of the tunneling probability and the initial position or momentum spread is capable of excluding the classical model.
For example, at $\sigma = 10$ fm the classical probability is more than three orders of magnitude smaller than the quantum one, which is at the measurable level of $10^{-3}$.

\subsection{Validity of the WKB approximation}
\label{sec:qtunneling}

Quantum tunneling is of course a well-studied phenomenon, with many interesting applications even in astrophysics~\cite{Nature-106.14,ZP-54.656}.
Traditionally, the projectile is assumed as an incident plane wave and the tunneling probability is approximated within the time-independent theory. One such treatment is the Wentzel-Kramers-Brillouin (WKB) approximation~\cite{ZP-39.828,ZP-38.518}.

\begin{figure}[H]
\centering
\includegraphics[width=\linewidth]{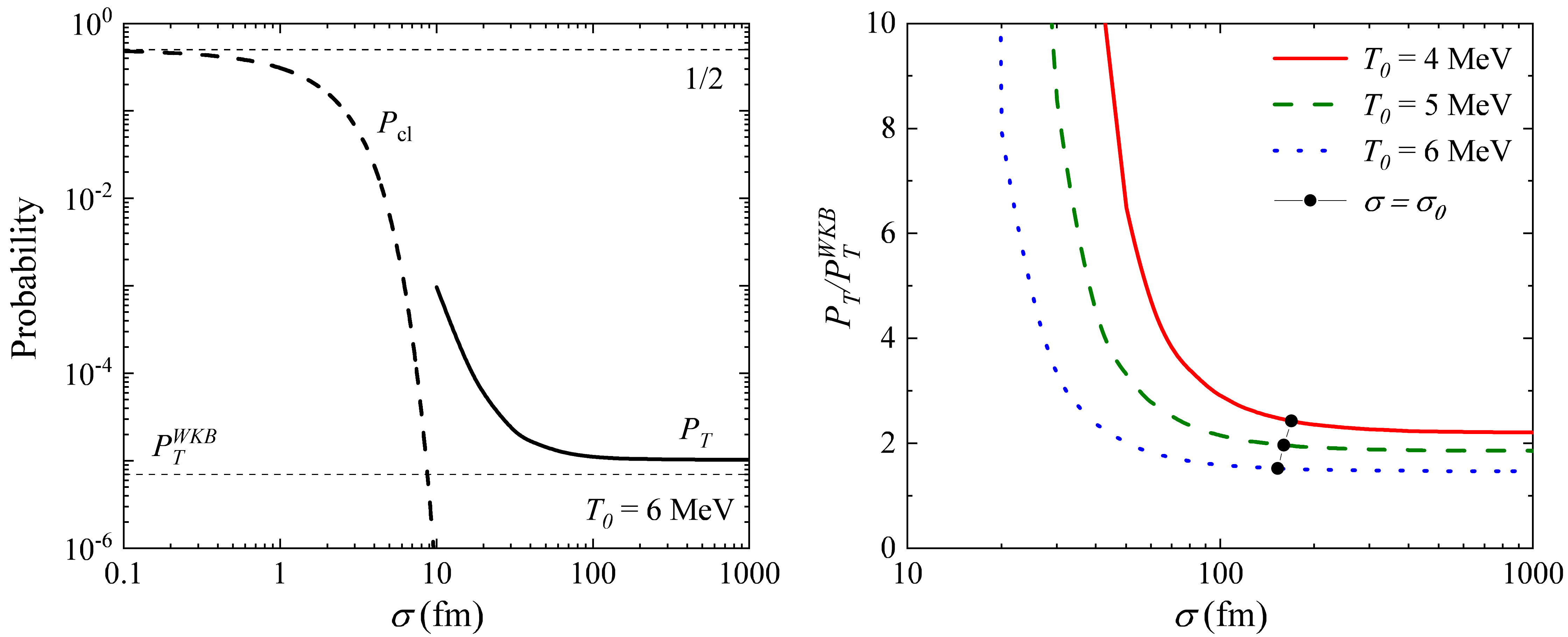}
\caption{Tunneling of alpha particle launched towards a Coulomb barrier of a gold nucleus. $\sigma$ is the spread of the initial position probability density of the projectile. $P_T$ is the tunneling probability obtained in the simulations of quantum dynamics. $P_T^\text{\emph{WKB}}$ is the WKB approximation of the same. $P_\text{cl}$ is the probability of barrier crossing obtained in the classical model (see main text). The initial launch distance is $L = 50$ pm. Numerical details are given in Appendix~\ref{appendix:constants}.}
\label{fig:TunnProb_50K}
\end{figure}

We now provide conditions under which this approximation matches the time-dependent approach presented here. Some advancements have already been made in the limiting case of low-energy projectiles with a negligible momentum variance~\cite{PLA-378.1071}, and it is expected that many peculiar effects arise with a large momentum spread~\cite{PR-40.621,PLA-220.41,PLA-225.303,JPA-37.2423,PLA-378.1071}.
In the left panel of Fig.~\ref{fig:TunnProb_50K} we plotted the tunneling probability in the WKB approximation by the horizontal dashed line at the bottom. The WKB result tends to the dynamical one in the limit of small initial momentum spread (large position spread).
However, it does not exactly approach the tunneling probability obtained in the dynamical simulation.
The right panel of Fig.~\ref{fig:TunnProb_50K} compares the two results for different kinetic energies. One can draw a boundary between the range where the two results are comparable and where they are very different.
It turns out that it is given by the optimal $\sigma_0$ we have derived above.

\section{Prospects and limitations}

An interesting regime of Rutherford scattering is when the projectiles are moving at relativistic speeds, e.g., the photons moving through a medium with refractive index $\sim 1/x$~\cite{AJP-81.405}.
While the non-relativistic wave packets shown in Fig.~\ref{fig:FindingSig0_5MeV} of the present work do not recover the classical solution, in the Rutherford scattering of photons the classical limit is achieved~\cite{AJP-81.405}. This is a consequence of the non-dispersive behaviour of photonic systems. For non-relativistic particles the variance in momentum directly translates to a variance in the velocity, which leads to a growing position spread in time.
Unlike alpha particles, photonic systems are non-dispersive as the variance in momentum does not correspond to a variance in the velocity~\cite{Ehrenfest-NicolasWheeler}. Accordingly, apart from the negligible impact of refractive index, there is no change in the position spread of a photonic wave function. One can therefore have an extremely narrow wave packet, and hence particle-like dynamics, all along the scattering event. This explains the recovery of classical solutions in the photonic Rutherford scattering~\cite{AJP-81.405}.

The time-dependent scattering problems, like the one discussed in this work, involve the notions of positions and trajectories. While one can talk about the expected positions for a wave packet, no such analogous counterpart exists for individual plane waves~\cite{ActaPhyPolB-33.2059}. Hence a consistent theory can only be built on the dynamics of wave functions with finite position spreads~\cite{AJP-81.405,ActaPhyPolA-101.369}.

In this work we represented the projectile as an incident Gaussian wave packet and the target with a simplistic Coulomb potential. Even though this works well in the considered energy range, this is an approximation on a few fronts. For one thing the target should also be considered as a wave packet and the scattering must be understood from the dynamics of two colliding wave packets~\cite{AJP-68.1113}. Secondly the internal structures of the projectile and the target will give rise to various exchange effects~\cite{NP-78.409,NPA-148.529,PR-155.29,PRC-61.054610}. These effects will be more prominent at higher energies and must be accounted for by an inclusion of exchange terms in the Hamiltonian~\cite{NP-47.652,PTP-122.1055,PLB-772.1}. An even fuller approach would be to replace the Coulomb potential with more realistic interactions~\cite{JPCC-122.14606,MicMach-11.319,JPCA-119.6897}.

Finally, if we consider a two or three-dimensional setting, different parts of the wave packet will acquire different phases as time passes. This will give rise to a diffraction pattern in the angular distribution of the total amplitude~\cite{AJP-52.60}. An extension of our work in two-dimensions promises to calculate the possible dependence of this pattern on the initial spread of the projectile wave function.

\section{Conclusions}

Average quantum dynamics in potentials with well-defined convexity properties does not approach its classical counterpart.
We have shown this for the case of Coulomb interaction between the alpha particles and the stationary gold nuclei in a head-on collision. The differences include the distance of closest approach, time of collision, and time symmetry of the dynamics.
We sketched an experiment aimed at verifying these predictions.
It could be rather challenging as we focused on head-on collisions in this work.
It would be interesting to work out possible differences in the 2D setting and compute how the differential cross-sections depend on the spread of the initial wave function of the alpha particle.
Such predictions might be easier to verify in a laboratory.
Finally, the model could be extended to study fully quantum mechanically astrophysical nuclear reactions.

\section*{Acknowledgements}

This work is jointly supported by \textbf{(i)} Nanyang Technological University, Singapore via NTU-India Connect Research Internship, \textbf{(ii)} Polish National Agency for Academic Exchange NAWA Project No. PPN/PPO/2018/1/00007/U/00001, and \textbf{(iii)} Indian Institute of Technology Roorkee, India via TSS-IRI grant. T.K. thanks Timothy Liew for hospitality at Nanyang Technological University, Singapore. The authors especially thank one of the anonymous reviewers for valuable comments and suggestions that have improved the presentation of this article.

\bibliographystyle{apsrev4-2}
\bibliography{refer_ruthexp}

\newpage
\appendixpage
\appendix
\begin{appendices}

\section{Cayley's form of evolution operator}
\label{appendix:cayley}

The Cayley's form of time evolution operator is given by~\cite{JCP-68.2794,JCP-134.041101,book_CompPhys_FJVesely,CPC-208.9,AJP-35.177}
\begin{equation}
U(t+\Delta t,t) \approx \left( 1 + i\frac{ H\Delta t}{2\hbar} \right)^{-1} \left( 1 - i\frac{ H\Delta t}{2\hbar} \right),
\end{equation}
which implies that $\psi(x,t)$ and $\psi(x,t+\Delta t)$ are related by an Implicit-Explicit Crank-Nicolson relation~\cite{book_CompPhys_FJVesely}:
\begin{equation}
\left( 1 + i\frac{ H\Delta t}{2\hbar} \right) \psi(x,t+\Delta t) = \left( 1 - i\frac{ H\Delta t}{2\hbar} \right) \psi(x,t).
\label{ForBackTimeEqn}
\end{equation}
Therefore, the idea is to evolve $\psi(x,t)$ by half of the time-step forward-in-time and $\psi(x,t+\Delta t)$ by half of the time-step backward-in-time, such that they both agree at time $t+\Delta t/2$. We approximate the second-order derivatives in Eq.~\eqref{ForBackTimeEqn} by the three point central difference formula to arrive at
\begin{equation}
\begin{split}
\psi_j^{n+1} + \frac{i \Delta t}{2\hbar} \left[ -\frac{\hbar^2}{2m} \left( \frac{\psi_{j+1}^{n+1} - 2\psi_{j}^{n+1} + \psi_{j-1}^{n+1}}{\Delta x^2} \right) + V_j  \psi_j^{n+1} \right] \\ = \psi_j^{n} - \frac{i \Delta t}{2\hbar} \left[ -\frac{\hbar^2}{2m} \left( \frac{\psi_{j+1}^{n} - 2\psi_{j}^{n} + \psi_{j-1}^{n}}{\Delta x^2} \right) + V_j  \psi_j^{n} \right],
\end{split}
\label{TLSE_eqn}
\end{equation}
where $\psi_j^n \equiv \psi(x_j,t_n)$, $\Delta x=x_{j+1}-x_j$ is the grid size, and $\Delta t=t_{n+1}-t_n$ is the time step. This is re-written in form of a matrix equation:
\begin{equation}
\begin{pmatrix}
b_1 & a \\ \ddots & \ddots & \ddots \\ & a & b_{j-1} & a \\ & & a & b_j & a \\ & & & a & b_{j+1} & a \\ & & & & \ddots & \ddots & \ddots \\ & & & & & a & b_{J-1}
\end{pmatrix} \\
\cdot \begin{pmatrix}
\psi_1^{n+1} \\ \vdots \\ \psi_{j-1}^{n+1} \\ \psi_{j}^{n+1} \\ \psi_{j+1}^{n+1} \\ \vdots\\ \psi_{J-1}^{n+1} \end{pmatrix} = \begin{pmatrix}
\zeta_1^{n} \\ \vdots \\ \zeta_{j-1}^{n} \\ \zeta_{j}^{n} \\ \zeta_{j+1}^{n} \\ \vdots \\ \zeta_{J-1}^{n}
\end{pmatrix},
\label{TLSE_matrix}
\end{equation}
where
\begin{equation}
a = -\frac{\hbar}{2m} \left( \frac{i \Delta t}{2 \Delta x^2} \right),
\end{equation}
\begin{equation}
\hspace{5mm} b_j = 1 + \frac{i \Delta t}{2\hbar} \left[ \frac{\hbar^2}{2m} \left( \frac{2}{\Delta x^2} \right) + V_j \right], \hspace{5mm} \text{and}
\end{equation}
\begin{equation}
\zeta_j^n = \psi_j^{n} - \frac{i \Delta t}{2\hbar} \left[ -\frac{\hbar^2}{2m} \left( \frac{\psi_{j+1}^{n} - 2\psi_{j}^{n} + \psi_{j-1}^{n}}{\Delta x^2} \right) + V_j  \psi_j^{n} \right].
\end{equation}
Eq.~\eqref{TLSE_matrix} now represents a tridiagonal system of linear equations for $J-1$ unknown wave function values at time $t+\Delta t$. This is solved for $\psi^{n+1}$ by performing a LU-factorisation of the tridiagonal matrix on the left, followed by forward- and backward substitutions of the $\zeta$ vector on the right.

The issue with the dynamics of long-range interactions is that they require huge computational resources. Moreover, the intrinsic time scales are much smaller than the time elapsed in the collision events. This makes it challenging to study the scattering in a fully quantum mechanical time-dependent theory~\cite{FBS-56.727,PRC-73.054608,CMAME-193.1733}. Within the scope of this work, the wave function remains confined near a single point; this allows us to perform a dynamic allocation of the grid and recast the problem as a sequential chain of small simulations. We start by defining a box around the initial launch distance such that the wave packet is highly concentrated at its center. The simulation is initiated, and the wave packet evolves in the Coulomb potential. We keep track of the peak and the width of the probability density function to make sure there is no reflection from the boundaries of the box. Once the tail of the probability distribution starts approaching either of the boundaries, we define a new box that re-confines the wave packet at its center. The final wave function from the old grid (box) is interpolated and used as an initial condition to start the calculation onto the new grid. The grid allocation and interpolation keep repeating until the wave packet returns at its original launch distance after the collision. This chain of small simulations is verified to precisely simulate the whole of the scattering event.

\section{Numerical details} \label{appendix:constants}

\begin{itemize}[leftmargin=*]

\item Numerical calculations are performed in natural units of $c=1$. Accordingly, the conversion constant $\hbar c = 197.3269804$ MeV fm. We follow these units within this section.

\item Fine-structure constant, $\alpha$ = $1/137.035999084$.

\item Mass of alpha particle, m = $3727.3794066$ MeV.

\item In Fig.~\ref{fig:10K5MeV}: $L = 10$ pm and $T_0 = 5$ MeV, which implies $P = 193.06$ MeV.

\item In Fig.~\ref{fig:FindingSig0_5MeV}: $T_0 = 5$ MeV implying $P = 193.06$ MeV; $\sigma_0$ = $71.49$, $101.10$, $123.82$, $142.97$, and $159.85$ fm for $L$ = $10$, $20$, $30$, $40$, and $50$ pm, respectively.

\item In Fig.~\ref{fig:qdoca}:
\begin{itemize}
\item $T_0 = 2$ MeV implies $P = 122.10$ MeV; $\sigma_0$ = $89.89$, $127.12$, $155.69$, $179.78$, and $201.00$ fm for $L$ = $10$, $20$, $30$, $40$, and $50$ pm, respectively.
\item $T_0 = 3$ MeV implies $P = 149.55$ MeV; $\sigma_0$ = $81.22$, $114.87$, $140.69$, $162.45$, and $181.62$ fm for $L$ = $10$, $20$, $30$, $40$, and $50$ pm, respectively.
\item $T_0 = 4$ MeV implies $P = 172.68$ MeV; $\sigma_0$ = $75.59$, $106.90$, $130.92$, $151.18$, and $169.02$ fm for $L$ = $10$, $20$, $30$, $40$, and $50$ pm, respectively.
\item $T_0 = 5$ MeV implies $P = 193.06$ MeV; $\sigma_0$ = $71.49$, $101.10$, $123.82$, $142.97$, and $159.85$ fm for $L$ = $10$, $20$, $30$, $40$, and $50$ pm, respectively.
\end{itemize}

\item In Fig.~\ref{fig:MicroProp_50K5MeV}: $L = 50$ pm and $T_0 = 5$ MeV, which implies $P = 193.06$ MeV and $\sigma_0 = 159.85$ fm.

\item In Fig.~\ref{fig:TunnProb_50K}: $L = 50$ pm, which implies $\sigma_0$ = $169.02$, $159.85$, and $152.73$ fm for $T_0$ = $4$, $5$, and $6$ MeV, respectively.

\item We have employed two different convergence tests to perform the error analysis: (i) the convergence of the quantum distance of closest approach, and (ii) the constancy of the expected Hamiltonian all along the scattering event. In Fig.~\ref{fig:convg-10K5MeV} we present the error analysis assuming $L=10$ pm and $\sigma = 100$ fm (the typical width of Gaussian wave packets considered in this work).

\begin{figure}[H]
\centering

\begin{subfigure}{0.9\textwidth}
\centering
\includegraphics[width=\linewidth]{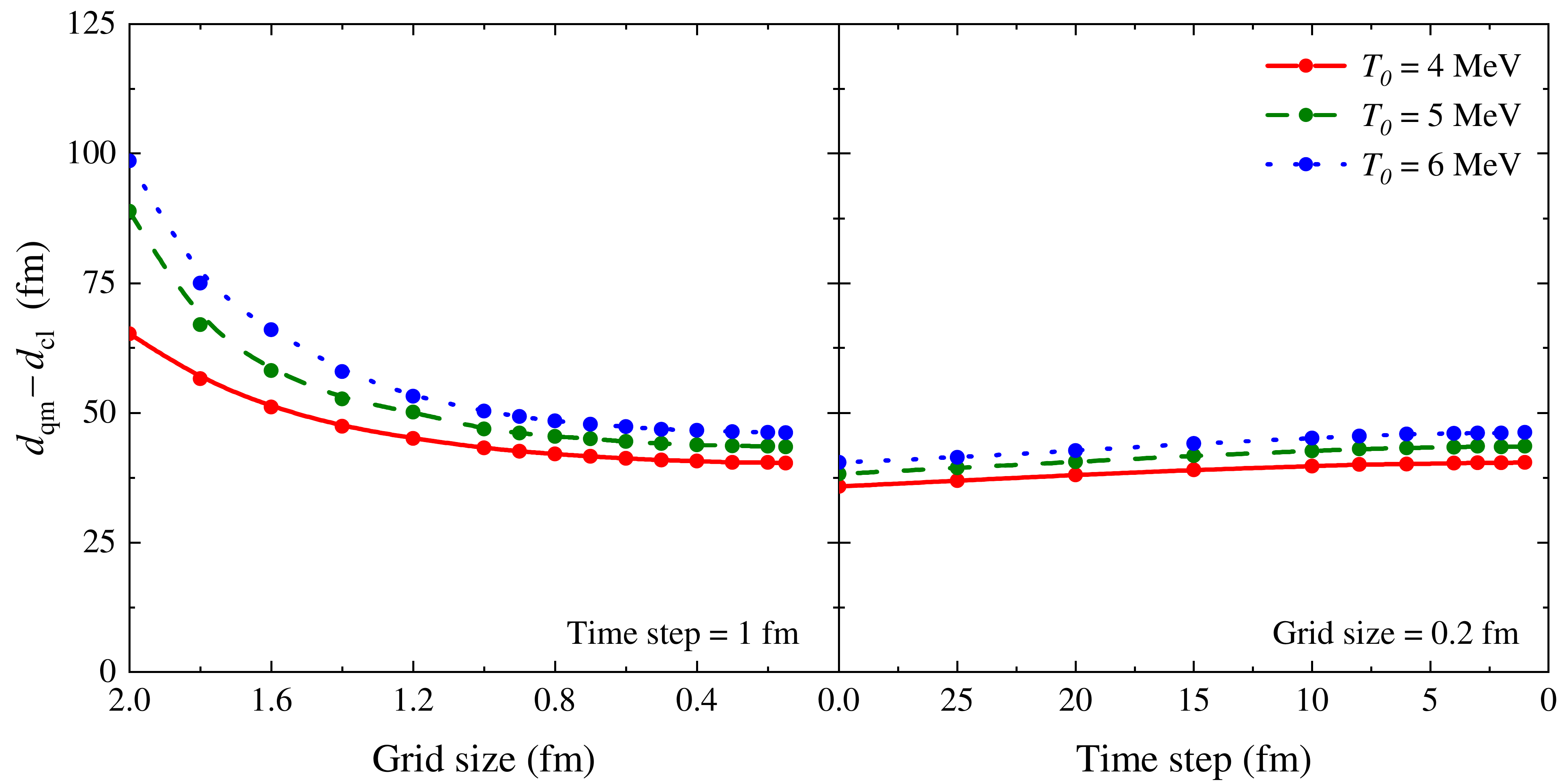}
\caption{The first test: convergence of the quantum distance of closest approach.}
\label{fig:convg_10K5MeV_dqm}
\end{subfigure}

\vspace{5mm}

\begin{subfigure}{\textwidth}
\centering
\includegraphics[width=\linewidth]{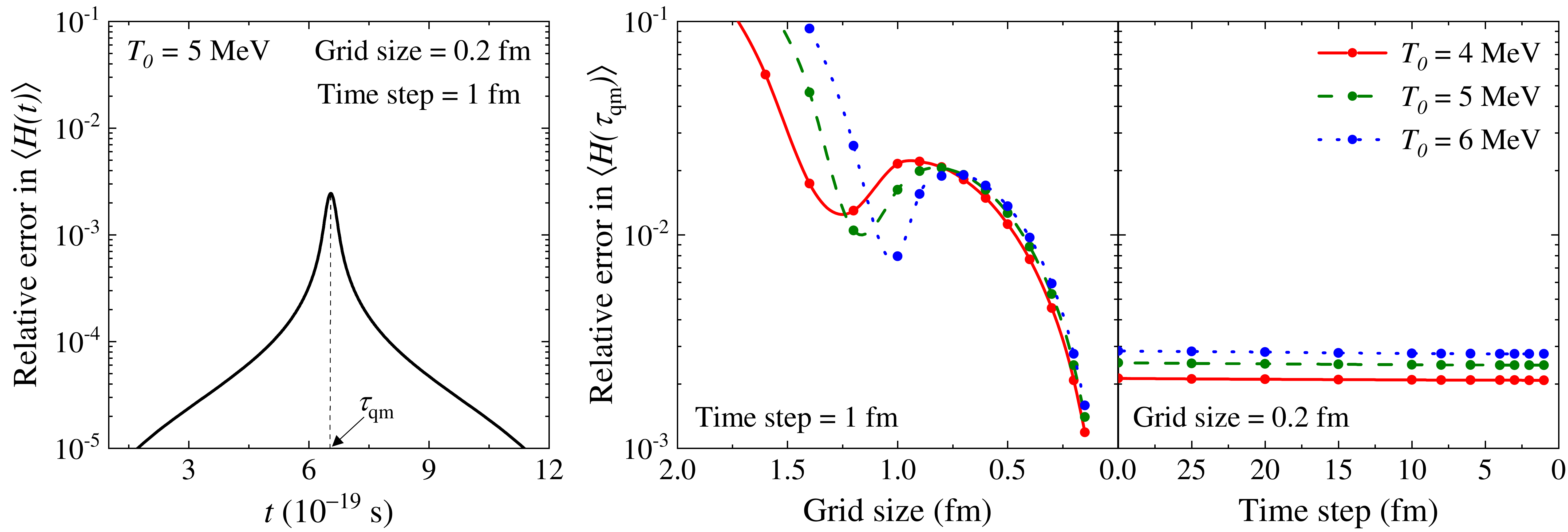}
\caption{The second test: constancy of the expected Hamiltonian during the scattering event. The relative error in $\ev{H(t)}$ is given by $1-\ev{H(t)}/\ev{H(0)}$.}
\label{fig:convg_10K5MeV_energy}
\end{subfigure}

\caption{Convergence of numerical results. The alpha particle Gaussian wave packet is initially centered at a separation $L=10$ pm with a width of $\sigma = 100$ fm. $T_0$ denotes the kinetic energy of an equivalent classical projectile. $d_\text{qm}$ and $d_\text{cl}$ represent the quantum and classical distances of closest approach, and $H$ is the Hamiltonian.}
\label{fig:convg-10K5MeV}

\end{figure}

Fig.~\ref{fig:convg_10K5MeV_dqm} shows that there is a very good convergence in $d_\text{qm}$ for grid size $\lesssim 0.3$ fm and time step $\lesssim 2$ fm. The first panel of Fig.~\ref{fig:convg_10K5MeV_energy} shows the variation of the relative error in expected Hamiltonian, i.e. $1-\ev{H(t)}/\ev{H(0)}$, as a function of time for a fixed grid size and time step. For most times this error is negligible, except at $t=\tau_{qm}$ where it attains its peak. The second panel shows this peak error is negligible for grid size $\lesssim 0.3$ fm. The last panel shows that an increased time step has no significant impact on the error in energy. Since this was not the case for the error in $d_\text{qm}$ (see the second panel in Fig.~\ref{fig:convg_10K5MeV_dqm}), we use two tests of convergence for reliable calculations.

\item A grid size of $0.2$ fm, with a time step of $1$ fm, ensures a good precision in all of our calculations. We have accordingly set the same throughout this work.

\end{itemize}

\end{appendices}

\end{document}